\documentclass{PoS}
\usepackage{amsmath}
\usepackage{amssymb,amsthm,amsfonts,bbm,dsfont,pifont}
\usepackage{slashed}
\usepackage{graphicx}

\newcommand{\PZ}{\text{Z}}
\newcommand{\PH}{\text{H}}
\newcommand{\PW}{\text{W}}

\newcommand{\rT}{\mathrm{T}}
\newcommand{\cut}{\mathrm{cut}}

\title{Electroweak accuracy in V-pair production \\  
                              at the LHC}
\ShortTitle{Electroweak accuracy in V-pair production at the LHC}

\author{Anastasiya Bierweiler\\
        Karlsruhe Institute of Technology (KIT), 
        Institut f\"ur Theoretische Teilchenphysik,\\
        D-76128 Karlsruhe, Germany \\ 
        E-mail: \email{nastya@particle.uni-karlsruhe.de}}
\author{\speaker{Tobias Kasprzik}\\
        Karlsruhe Institute of Technology (KIT), 
        Institut f\"ur Theoretische Teilchenphysik, \\        
        D-76128 Karlsruhe, Germany\\
        E-mail: \email{kasprzik@particle.uni-karlsruhe.de}}
\author{Johann H. K\"uhn \\ 
        Karlsruhe Institute of Technology (KIT), 
        Institut f\"ur Theoretische Teilchenphysik,\\
        D-76128 Karlsruhe, Germany \\ 
        E-mail: \email{jk@particle.uni-karlsruhe.de}}

      \abstract{ Vector-boson pair production is of great
        phenomenological importance at the LHC. These processes will
        help to validate the Standard Model at highest energies, and
        they may also open the door for the discovery of new physics
        potentially showing up in subtle modifications of the
        non-abelian structure of weak interactions. In this letter, we
        present the first full $\mathcal{O}(\alpha^3)$ analysis of
        on-shell W$^\pm$Z and Z-pair production at the LHC with all mass
        effects consistently included. The resulting electroweak
        corrections are negative,  strongly increase with increasing
          transverse momenta, and lead to significant modifications of
          rapidity and angular distributions. In view of the high
        energies accessible at the LHC, combined with considerable
          event rates, our results have to be included in a proper
        analysis of experimental
        data. \\
\begin{flushright}
\emph{TTP12-028 \\ 
LPN12-089 \\ 
SFB/CPP-12-59}
\end{flushright}}

\FullConference{36th International Conference of High Energy Physics - ICHEP2012,\\
		July 4-11, 2012\\
		Melbourne, Australia}

\begin{document}
\section{Introduction}
A profound understanding of vector-boson pair production processes at
the LHC is desirable for various reasons. Such processes not only
contribute an important irreducible background to Standard-Model (SM)
Higgs production at moderate energies, but will also provide deeper
insight into the physics of the weak interaction at the high-energy
frontier, possibly even allowing for the discovery of BSM
physics. Consequently, great effort has been made during the last
years to push the theory predictions for this process class to a new
level, where, besides the dominating QCD corrections, also electroweak
(EW) effects have been studied extensively (see, e.g.,
Ref.~\cite{Bierweiler:2012} and references therein).

Extending our work on the EW effects in W-pair production at the
LHC~\cite{Bierweiler:2012}, we present corresponding results for
on-shell W$^\pm$Z and ZZ pair production in the SM.  We will restrict
ourselves to pair production through quark--antiquark
annihilation. Photon--photon collisions do not contribute to WZ
production (in contrast to the case of W pairs), are of higher order for
Z-pair production and will not be discussed further. Also gluon fusion
is, evidently, irrelevant for WZ production. For the case of W-pair
production we have demonstrated that gluon fusion amounts to order of
5\% relative to the quark--antiquark annihilation with decreasing
importance for increasing transverse momenta. Gluon fusion, furthermore,
does not lead to a strong modification of the angular and rapidity
distributions of the W bosons. A qualitatively similar behaviour is
expected for ZZ production through gluon fusion, which therefore will
not be discussed further. Also QCD corrections for WZ and ZZ production
are expected to be similar to those for W pairs discussed in
Ref.~\cite{Bierweiler:2012} and will not be analyzed in the present
paper, which, instead, will be entirely devoted to genuine EW
corrections. Earlier papers for gauge-boson pair production have
emphasized the drastic influence of the EW corrections on cross sections
and distributions in the high-energy region~\cite{Accomando:2001fn,
  Accomando:2004de, Accomando:2005ra}, neglecting at the same time terms
of order $M_{\PW}^2/\hat{s}$.  In the present paper, the full mass
dependence, namely terms proportional to powers of $M_V^2/\hat{s},$ is
consistently accounted for to obtain results valid in the whole energy
range probed by LHC experiments.  Therefore, our results are
complementary to those presented in Ref.~\cite{Accomando:2004de}, where
only logarithmic corrections were considered, but the leptonic decays of
the vector bosons and related off-shell effects were included in a
double-pole approximation. Comparing both approaches, we try to estimate
the remaining theoretical uncertainties related to EW corrections in
this important process class.

\section{Details of the calculation}
At the LHC, WW, W$^\pm$Z and ZZ production
is, at lowest order $\mathcal{O}(\alpha^2)$, induced by the partonic processes
\begin{subequations}
\label{eq:LO}
\begin{eqnarray}
  q\;\bar{q} &\to& \mathrm{W^-W^+} \quad (q = \mathrm{u,d,s,c,b})\,, \\
  \mathrm{u}_i\mathrm{\bar{d}}_j &\to& \mathrm{W^+Z}\,, \quad \mathrm{\bar{u}}_i\mathrm{d}_j \to \mathrm{W^-Z} \quad (i = 1,2;\;
  j=1,2,3)\,, \\
  q\;\bar{q} &\to& \mathrm{ZZ}\,,
\end{eqnarray}
\end{subequations}
where the hadronic results are obtained by convoluting the partonic
cross sections with appropriately chosen PDFs and summing incoherently
over all contributing channels. To allow for consistent predictions with
full $\mathcal{O}(\alpha^3)$ accuracy, virtual EW corrections, as well
as real corrections due to photon radiation have to be considered. The
evaluation of the radiative corrections is based on the well-established
{\tt Feyn\-Arts/FormCalc/LoopTools} setup~\cite{Kublbeck:1990xc, Hahn:2000kx,
  Hahn:1998yk, Hahn:2001rv, vanOldenborgh:1989wn},  the WW process has
been independently cross-checked by a setup based on
{\tt QGraf}~\cite{Nogueira:1991ex} and {\tt Form}~\cite{Vermaseren:2000nd}.  To
considerably reduce the computational effort, light quark masses are
neglected whenever possible. However, soft and collinear singularities
occurring in intermediate steps of the calculation are regularized by
small quark masses $m_q$ and an infinitesimal photon mass $\lambda$,
generating unphysical $\ln m_q$ and $\ln \lambda$ terms. To allow for a
numerically-stable evaluation of those infrared-divergent parts of the
cross sections related to real radiation, the phase-space slicing method
is adopted. Finally, adding real and virtual contributions, the
regulator-mass dependence drops out in any properly defined physical
result. The input parameters to be specified in Section~\ref{se:setup}
are renormalized in a modified on-shell scheme~\cite{Denner:1991kt},
where the Fermi constant $G_\mu$ is used instead of $\alpha(0)$ to
effectively account for universal corrections induced by the running of
$\alpha(\mu)$ to the weak scale~\cite{Dittmaier:2001ay}.

\section{Input and setup}
\label{se:setup}
For the computation presented here the same setup as specified in Ref.~\cite{Bierweiler:2012} is
applied. We use the following SM input parameters for the numerical
analysis,
\begin{equation}\arraycolsep 2pt
\begin{array}[b]{lcllcllcl}
G_{\mu} & = & 1.16637 \times 10^{-5} \;\mathrm{GeV}^{-2}, \quad & & & & &  \\
M_{\mathrm{W}} & = & 80.398\;\mathrm{GeV}, & M_{\mathrm{Z}} &
= & 91.1876\;\mathrm{GeV}, \quad & & & \\ 
 M_\PH & = & 125\;\mathrm{GeV}, & m_{{\rm t}} & = & 173.4\;\mathrm{GeV}\,, & & & 
\end{array}
\label{eq:SMpar}
\end{equation}
which are taken from Ref.~\cite{Amsler:2008zzb}. In the on-shell scheme
applied in our computation,
the weak mixing angle $\cos^2\theta_{\mathrm{w}} =
M_{\mathrm{W}}^2/M_{\mathrm{Z}}^2$ is a derived quantity.
For the computation of the processes~\eqref{eq:LO} and the corresponding EW
radiative corrections, we use the MSTW\-2008\-LO PDF
set~\cite{Martin:2009iq} in the LHAPDF setup~\cite{Whalley:2005nh}. In
order to consistently include $\mathcal{O}(\alpha)$ corrections, in
particular real radiation with the resulting collinear singularities,
PDFs in principle should take these QED effects into account. Such a PDF
analysis has been performed in Ref.~\cite{Martin:2004dh}, and the
$\mathcal{O}(\alpha)$ effects are known to be small, as far as their
effect on the quark distribution is concerned~\cite{Roth:2004ti}. In
addition, the currently available PDFs incorporating
$\mathcal{O}(\alpha)$ corrections~\cite{Martin:2004dh} include QCD
effects at NLO, whereas our EW analysis is LO with respect to
perturbative QCD only. For these reasons, the MSTW2008LO set is used as
our default choice for the quark-induced processes. The renormalization
and factorization scales are always identified, our default scale choice
being the phase-space dependent average of the vector-boson transverse masses
\begin{equation}
\mu_{{\rm R}} = \mu_{{\rm F}} = \overline{m_{\rT}} =
\frac{1}{2}\left(\sqrt{M_{V_1}^2+p_{\rT,V_1}^2}+\sqrt{M_{V_2}^2+p_{\rT,V_2}^2}\right)\,.
\end{equation}
A similar scale choice was taken in Ref.~\cite{Accomando:2004de} for the
computation of the EW corrections to four-lepton production at the LHC. 
In our default setup, we require a minimum transverse momentum and a maximum
rapidity for the final-state vector bosons,
\begin{equation}\label{eq:defcuts}
 p_{\rT,V_i} > 15\;\mathrm{GeV}\,,\quad |y_{V_i}|<2.5\,,
\end{equation}
to exclude events where the bosons are emitted collinearly to the
initial-state partons. 

\section{Numerical results}
\label{se:numres}
In this section we present integrated cross sections for WW, W$^\pm$Z
and ZZ production at the CERN LHC with center-of-mass (CM) energies of
$\sqrt{s} = 8$~TeV (LHC8) and $\sqrt{s} = 14$~TeV (LHC14),
respectively. Results for the corresponding full EW corrections to the
processes $\mathrm{pp} \to \mathrm{W^\pm Z}+X$ and $\mathrm{pp} \to
\mathrm{ZZ}+X$ are shown for the first time, while the contributions for
W-pair production have already been presented in
Ref.~\cite{Bierweiler:2012} and are listed only for comparison.  In the
following, relative corrections $\delta$ are defined through
$\sigma_{\mathrm{NLO}} = (1+\delta)\times\sigma_{\mathrm{LO}}$.

\begin{figure}[t!]
\includegraphics[width = 1.0\textwidth]{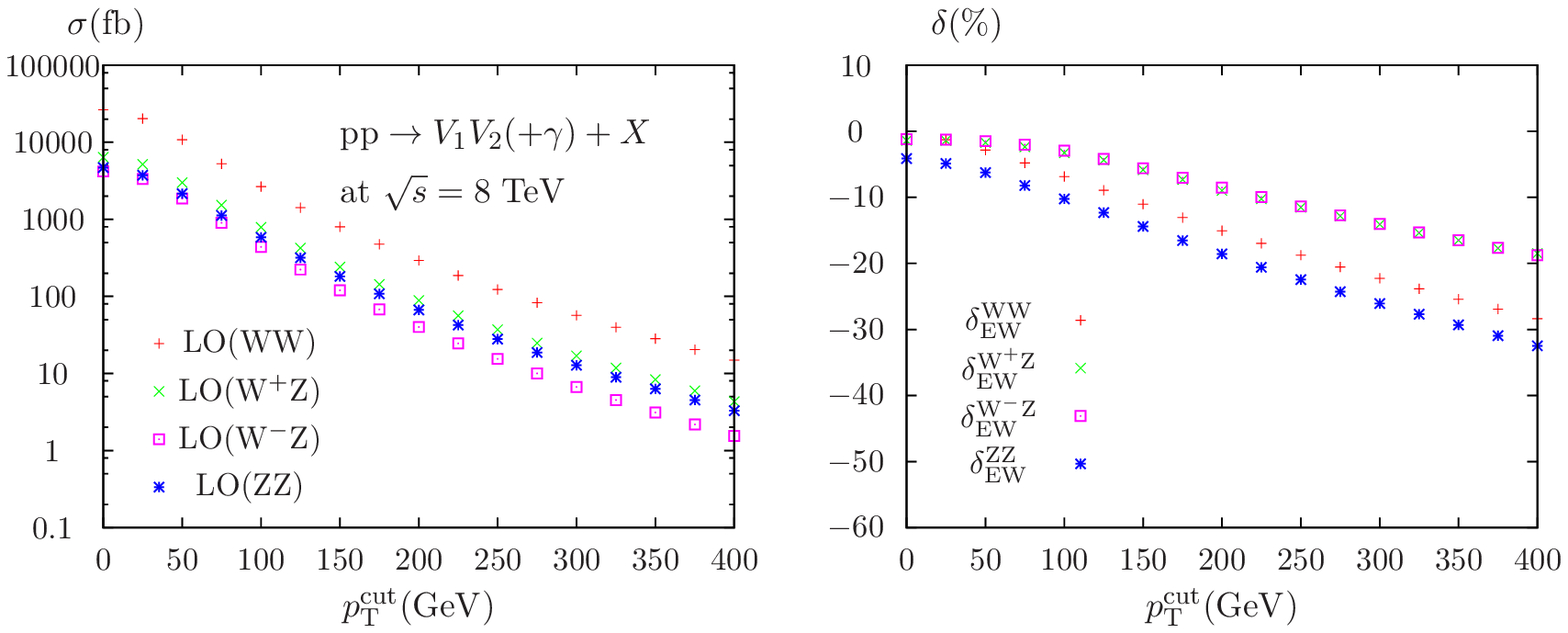}
\includegraphics[width = 1.0\textwidth]{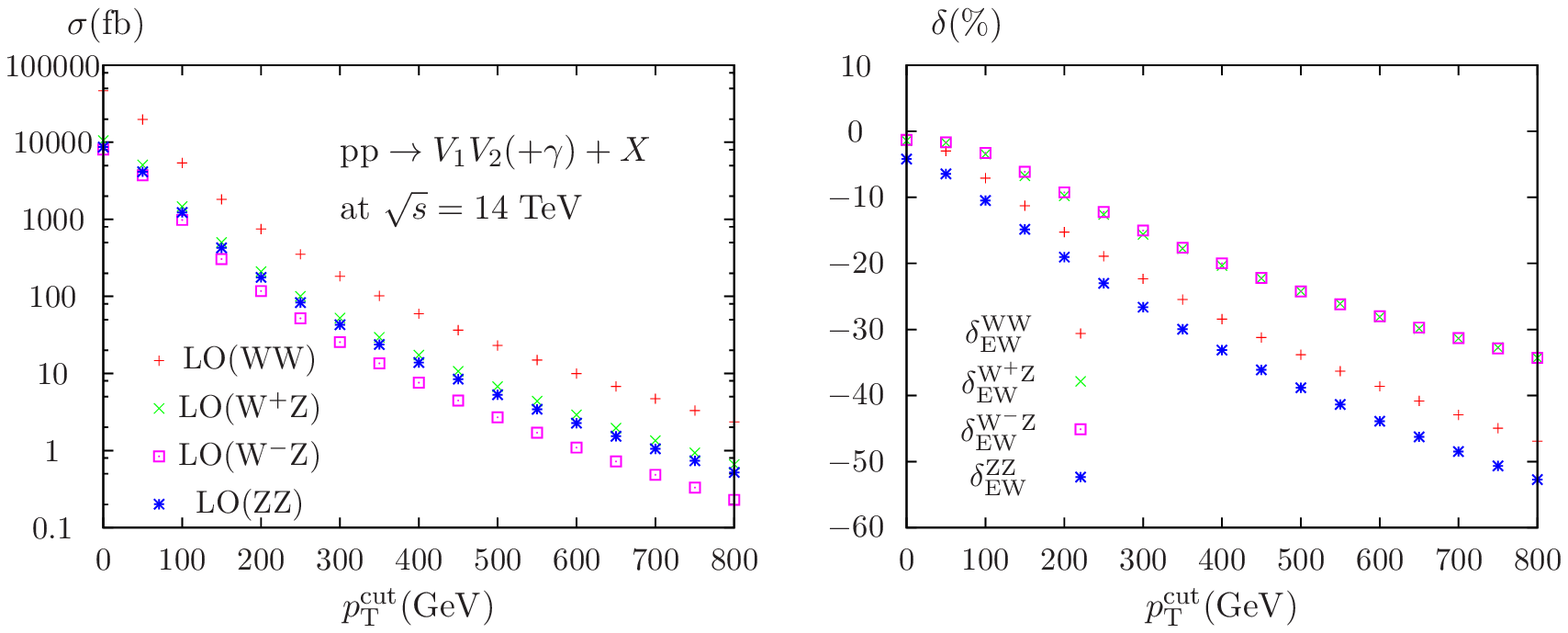}
\caption{\label{fi:ptcut}
Leading-order cross sections (left) and
  corresponding relative EW corrections (right) to WW, W$^\pm$Z and ZZ
  production at the LHC for different cuts on the vector-boson
  $p_{\rT}$.}
\end{figure}

Figure~\ref{fi:ptcut} shows integrated cross sections and the respective
EW corrections for different cut values on the boson transverse momenta
at LHC8 and LHC14, respectively, where the plot range has been adapted
to the CM energy available. The $p_{\rT}$-dependent relative EW
corrections are largest for ZZ production and similar to those in WW
production, for WZ production they are significantly smaller.  At LHC8,
for ZZ production corrections of $-30\%$ are observed for
\mbox{$p_{\rT}^{\cut} = 400$ GeV,} rising to $-50\%$ at $p_{\rT}^{\cut}
= 800$ GeV for $\sqrt{s} = 14$ TeV. As expected, the relative
corrections to W$^+$Z production coincide with those to W$^-$Z
production at the percent level as a consequence of the identical
corresponding unpolarized partonic cross sections.

\begin{figure}[t!]
\includegraphics[width = 1.0\textwidth]{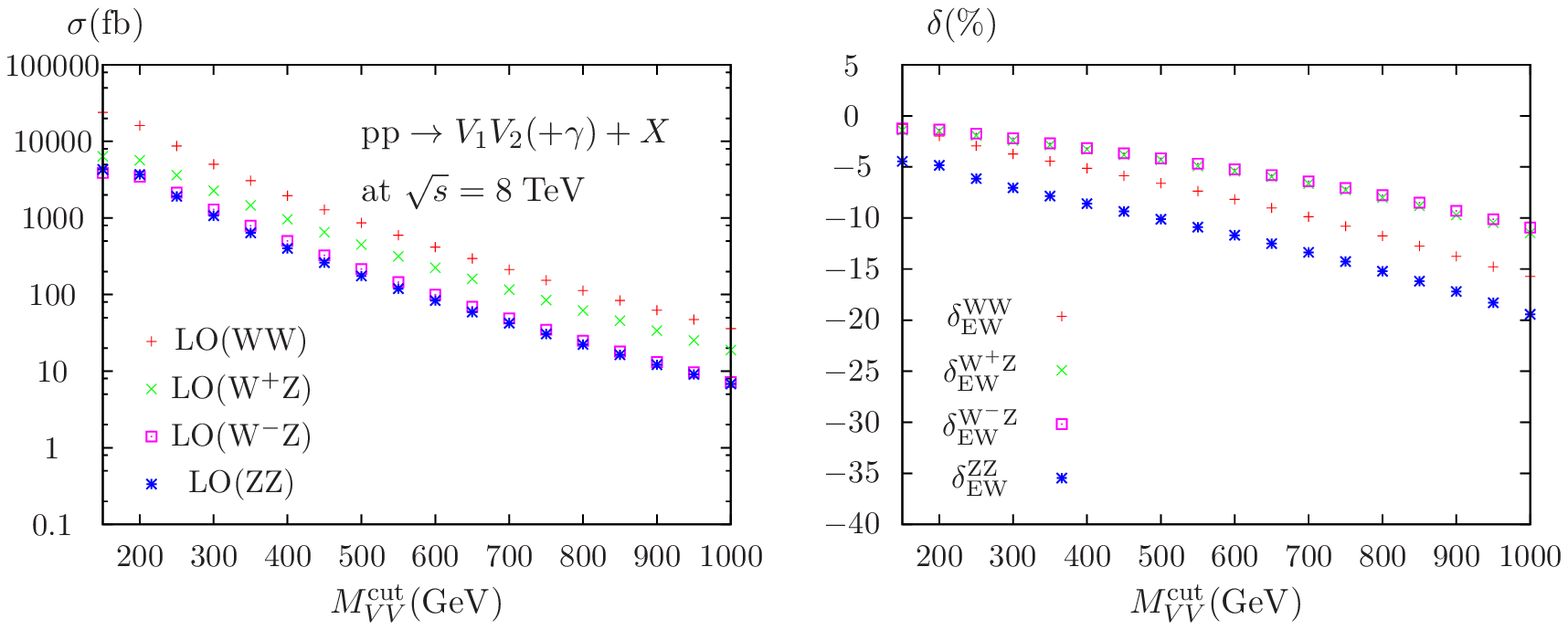}
\includegraphics[width = 1.0\textwidth]{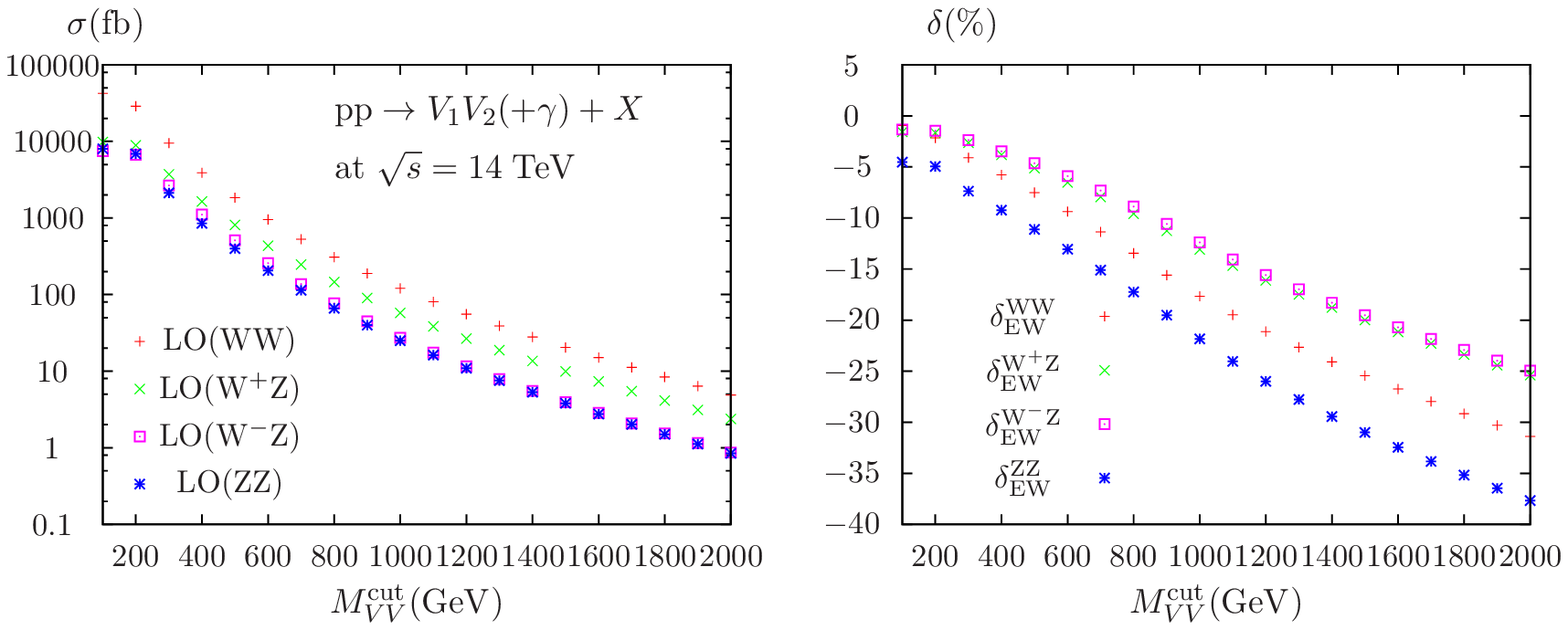}
\caption{\label{fi:mvvcut}
  Leading-order cross sections (left) and
  corresponding relative EW corrections (right) to WW, W$^\pm$Z and ZZ
  production at the LHC for different cuts on the vector-boson
  invariant mass.}
\end{figure}

In Figure~\ref{fi:mvvcut}, integrated cross sections are presented for
different values of the minimal invariant masses of the final-state
boson pair at LHC8 and LHC14, respectively. Again, one observes
negative EW corrections rising with the energy. However, the relative
corrections are smaller than in the $p_{\rT}^{\cut}$ scenario, since
vector-boson pair production at high invariant masses is dominated by
collinear emission at low $p_{\rT}$ (corresponding to small $\hat{t}$)
which is not affected by large negative Sudakov logarithms.

Finally, we compare our results ($\delta_{\mathrm{EW}}^{\mathrm{ZZ}}$)
to those given in Table 3 of Ref.~\cite{Accomando:2004de}
($\delta_{\mathrm{EW}}^{\mathrm{ZZ, ADK}}$), which were obtained in the
high-energy limit. We observe very good agreement in the whole energy
range considered if we employ the additional constraint $|\Delta
y_{\PZ\PZ}| < 3$ on the rapidity gap of the Z bosons to explicitly
enforce Sudakov kinematics (i.e.\ $\hat{s},\hat{t},\hat{u} \gg
M_{\PZ}^2$). For instance, we find $\delta_{\mathrm{EW}}^{\mathrm{ZZ}} =
-28.5\%$, to be compared with $\delta_{\mathrm{EW}}^{\mathrm{ZZ, ADK}} =
-28.1\%$ for $M_{\PZ\PZ}>1000$ GeV. As expected, EW corrections to ZZ
production in the Sudakov regime are exhaustively described by
logarithmic weak corrections, and mass effects do not play a significant
role. Surprisingly, also off-shell effects and final-state photon
radiation, both included in Ref.~\cite{Accomando:2004de}, as well as LHC
acceptance cuts, do not seem to noticeably affect the relative EW
corrections after the Z reconstruction has been performed.

\section{Conclusions}
\label{se:concl}
Extending our work on W-pair production at hadron colliders, we have
computed the full EW corrections to on-shell W$^\pm$Z and ZZ
production at the LHC. At high transverse momenta, the relative
corrections are found to be largest in the ZZ case and moderate for WZ
production. in case of ZZ production at the LHC we find perfect
agreement with older results obtained in the high-energy
approximation. In the future, a more detailed discussion of the
presented results will follow~\cite{Bierweiler:2012b}, including
predictions for differential cross sections.

\subsection*{Acknowledgements}
\noindent
This work has been supported by ``Strukturiertes Promotionskolleg
Elementarteilchen- und Astro\-teilchen\-physik'', SFB TR9 ``Computational
and Particle Physics'' and BMBF Contract 05HT4VKATI3.

\end{document}